\documentclass[%
 reprint,
superscriptaddress,
 amsmath,amssymb,
 aps,
 pra,
]{revtex4-2}

\usepackage{graphicx}
\usepackage{xcolor}
\usepackage{multirow}
\usepackage{caption}
\usepackage[version=4]{mhchem}
\usepackage[hidelinks]{hyperref}
\usepackage{listings}
\usepackage{dcolumn}
\usepackage{bm}
\usepackage{siunitx}

\begin{document}

\preprint{APS/123-QED}

\title{Quadratic Zeeman and Electric Quadrupole Shifts in Highly Charged Ions}

\author{Jan Gilles}
\affiliation{Fundamentale Physik für Metrologie FPM, Physikalisch-Technische Bundesanstalt PTB, D-38116 Braunschweig, Germany}
\affiliation{Institut für Mathematische Physik, Technische Universität Braunschweig, D-38116 Braunschweig, Germany}
\affiliation{Technische Universität München, D-80333 München, Germany}
\affiliation{Ludwig-Maximilians-Universität München, D-80539 München, Germany}
\author{Stephan Fritzsche}
\affiliation{Helmholtz-Institut Jena, D-07743 Jena, Germany}
\affiliation{GSI Helmholtzzentrum für Schwerionenforschung, D-64291 Darmstadt, Germany}
\affiliation{Theoretisch-Physikalisches Institut, Friedrich-Schiller-Universität, D-07743 Jena, Germany}
\author{Lukas J. Spieß}
\affiliation{Institute for Experimental Quantum Metrology QUEST, Physikalisch-Technische Bundesanstalt PTB, D-38116 Braunschweig, Germany}
\author{Piet O. Schmidt}
\affiliation{Institute for Experimental Quantum Metrology QUEST, Physikalisch-Technische Bundesanstalt PTB, D-38116 Braunschweig, Germany}
\affiliation{Institut für Quantenoptik, Leibniz Universität Hannover, D-30167 Hannover, Germany}
\author{Andrey Surzhykov}
\affiliation{Fundamentale Physik für Metrologie FPM, Physikalisch-Technische Bundesanstalt PTB, D-38116 Braunschweig, Germany}
\affiliation{Institut für Mathematische Physik, Technische Universität Braunschweig, D-38116 Braunschweig, Germany}

\date{\today}

\begin{abstract}
Recent advances in high-precision spectroscopy of highly charged ions necessitate an understanding of energy shifts of ionic levels caused by external electric and magnetic fields. Beyond the well-known Stark and linear Zeeman shifts, trapped ions may also exhibit quadratic Zeeman and electric quadrupole shifts. In this contribution, we present a systematic approach for the theoretical analysis of these shifts for arbitrary many-electron ions.
Based on the derived expressions and making use of the multiconfigurational Dirac-Fock approach, we performed calculations of quadratic Zeeman shift coefficients and quadrupole moments for various ionic states in \ce{Ca^{14+}}, \ce{Ni^{12+}} and \ce{Xe^{q+}} ions. These ions attract particular interest for ongoing and future experiments in optical clocks and tests of fundamental physics.
\end{abstract}

\maketitle

\section{\label{sec:introduction}Introduction}

Recent years have witnessed remarkable progress in the development of high-precision spectroscopy of highly charged ions (HCI). Innovative techniques such as sympathetic cooling and quantum logic spectroscopy \cite{cooling, quantum_logic} have enabled measurements of narrow transitions in HCI with frequency uncertainties on the order of $10^{-17}$ with the potential to reach $10^{-18}$ \cite{Ar13+, HCI}, which would be comparable to the presently best optical atomic clocks \cite{dimarcq_roadmap_2024}.
These measurements open up new avenues for high-precision tests of QED, investigations into nuclear properties and serve as a benchmark for advanced atomic structure theories. Moreover, HCI are emerging as promising candidates in the search for new physics beyond the Standard Model \cite{new_physics, HCI}. For instance, experiments on the isotope shift in HCI aim at the search for hypothetical fifth forces \cite{Ar13+, Highly_charged_Xe, isotope_shift}.

In order to obtain accurate results in HCI spectroscopy, comprehensive knowledge of all systematic uncertainties is required. Sources of these uncertainties arise due to the electric and magnetic fields present in the confining ion traps \cite{itano}. Typically, ions are stored in a Paul trap, where they are exposed to electric fields and a static magnetic bias field. In addition to the well-understood Stark and linear Zeeman shifts, these fields may also induce electric quadrupole and quadratic Zeeman shifts of atomic levels involved.
The electric quadrupole shift, for instance, affects atomic states with angular momentum $J > 1/2$, for which the deviation from spherical symmetry leads to a permanent electric quadrupole moment of the electron cloud. When subjected to an external electric field gradient, this leads to an energy shift of atomic levels, which depends on the orientation of the field relative to the quantisation axis as well as on the magnitude of the gradient. For HCI based optical clocks \cite{Ar13+}, the electric field gradient is dominantly caused by DC electric fields required for ion trapping in a linear Paul trap. A secondary contribution to the field gradient arises from the presence of logic ion required for quantum logic spectroscopy \cite{quantum_logic}.
The quadratic Zeeman shift is a second-order effect caused by the presence of an external homogeneous magnetic field. Specifically, such a field breaks the rotational symmetry of the atomic Hamiltonian, leading to level mixing. This mixing induces an additional magnetic dipole moment, causing an energy shift proportional to the square of the flux density.

During recent years, several theoretical and experimental studies have been reported on electric quadrupole and quadratic Zeeman shifts for neutral atoms and singly charged ions \cite{al+_qzsc, Sr_clock, Yb_clock, lange, dube_sr, Sr_theory, Sr+_EQS, Yb+_2D3/2, Yb+_2F7/2, Yb+_theory, Hg+_quadrupole_moment, Mg_Taichenachev, Mg_QZSC_chin, Sr_6}, and general scaling laws for reduced shift coefficients in HCI have been derived \cite{HCI_transitions}. However, to the best of our knowledge, there is little to no information available for \ce{Ca^{14+}}, \ce{Ni^{12+}} \cite{Ni12+} and \ce{Xe^{q+}}, $q \geq 9$ ions.
These ions are of special interest for future experiments as they might offer insights on hypothetical fifth forces \cite{isotope_shift, Highly_charged_Xe} as well as a proof of concept for HCI optical clocks \cite{Ni12+}.
Furthermore, only scattered theoretical details are available on the electric quadrupole and quadratic Zeeman shifts, and to the best of our knowledge, no explicit derivation from first principles can be readily found in the literature.

In this contribution, we lay down the general and universal theoretical background for the description of electric quadrupole and quadratic Zeeman shifts.
We start our analysis from the Hamiltonian of an arbitrary many-electron system exposed to an external electromagnetic field, as outlined in Section \ref{sec:Theoretical Background}.
Two perturbative terms in the Hamiltonian account for the interaction of an electron cloud with external electric and magnetic fields. In Section \ref{sec:Electric field coupling}, we address the term for electric field coupling, which is treated using a multipole expansion. This expansion naturally leads to a straightforward representation of Stark and electric quadrupole shifts. In Section \ref{sec:Magnetic field coupling}, we utilize the symmetric gauge for a vector potential of homogeneous magnetic fields, which simplifies calculations of magnetic field shifts within the framework of perturbation theory.
To numerically evaluate the electric quadrupole and quadratic Zeeman shifts derived in Section \ref{sec:Theoretical Background}, knowledge of atomic or ionic wave functions is required. The construction of such wave functions based on the multiconfigurational Dirac-Fock approach is presented in Section \ref{sec:Numerical Methods}.
Although the primary aim of this work is to investigate the electric quadrupole and second-order Zeeman shifts of particular HCI, Section \ref{sec:Results} starts with calculations for neutral and singly charged ions. These calculations serve to assess the reliability of the methods laid out in this work by comparison with existing data.
We then extend our analysis to shifts in \ce{Ca^{14+}}, \ce{Ni^{12+}} and \ce{Xe^{q+}}, $q=9,10,11,12,15,16,17$.
Finally, Section \ref{sec:Summary} provides a summary of our results and discusses potential directions for future research.

Unless explicitly stated otherwise, we will use Hartree atomic units: $\hbar = e = m_e = 4\pi\epsilon_0 = 1$. The Bohr magneton in these units is $\mu_B = 1/2$.

\section{\label{sec:theory}Theoretical Background}\label{sec:Theoretical Background}

In our work, we will employ a perturbative approach to study the quadratic Zeeman and electric quadrupole shifts of atomic levels. This analysis requires first to define the Hamiltonian of an unperturbed ion (or atom) in the absence of external electromagnetic fields. Within the relativistic framework, the unperturbed many-particle Hamiltonian reads as

\begin{equation}
    \mathcal{H}_0 = \sum_{i=1}^N \left( c \bm{\alpha}_i \bm{p}_i + \beta_i m c^2 + V_{nuc}(\bm{r}_i) + \sum_{j>i} V_{ee}(\bm{r}_i, \bm{r}_j) \right),
    \label{Dirac_Coulomb_Hamiltonian}
\end{equation}

where the summation runs over all bound electrons. In this expression, $\bm{\alpha}_i$ and $\bm{p}_i$ are the vectors of alpha matrices and momentum operators acting on the i-th electron and $\beta_i$ is the beta matrix. The scalar operators $V_{nuc}(\bm{r}_i)$ and $V_{ee}(\bm{r}_i, \bm{r}_j)$ describe the electron-nucleus and electron-electron interactions, respectively. In relativistic calculations $V_{ee}$ is usually given by the sum of the Coulomb term and the Breit correction, which accounts for first-order magnetic and retardation effects.

In general, generating any approximate solution of the eigenvalue equation $\mathcal{H}_0 |\psi \rangle = E |\psi \rangle$ for the Hamiltonian (\ref{Dirac_Coulomb_Hamiltonian}) is a rather complicated task, which requires the application of a suitable many-electron approach. In the present work, we will employ the Dirac-Fock (DF) and multiconfiguration Dirac-Fock (MCDF) approaches, which will be briefly discussed later. Both theories deal with eigenstates $|\Gamma J M \rangle$ of $\mathcal{H}_0$, characterized by their total angular momentum $J$ and its projection $M$. Furthermore, $\Gamma$ represents the set of all other quantum numbers necessary for a unique specification of the state.
In what follows, the eigensolutions $|\Gamma J M \rangle$ will be used as a basis for our perturbative calculations.

Having discussed the Hamiltonian of an unperturbed ion, we are now ready to investigate its interaction with external \textit{static} electric and magnetic fields. By using minimal coupling, the Hamiltonian of an ion exposed to these fields is given by

\begin{align}
\begin{split}
    \mathcal{H} &= \mathcal{H}_0 + \mathcal{H}_{el} + \mathcal{H}_{mag} \\
    &=\mathcal{H}_0 - \sum_{i=1}^N \phi(\bm{r}_i) + \sum_{i=1}^N c \bm{\alpha}_i \cdot \bm{A}(\bm{r}_i).
    \label{Hamiltonian_external_fields}
\end{split}
\end{align}

Here, the scalar and vector potentials $\phi$ and $\bm{A}$ describe static electric and magnetic fields, respectively. Both terms $\mathcal{H}_{el}$ and $\mathcal{H}_{mag}$ in (\ref{Hamiltonian_external_fields}) may lead to various shifts of ionic levels as can be obtained within the framework of perturbation theory. In what follows, we discuss the perturbative treatment of $\mathcal{H}_{el}$ and $\mathcal{H}_{mag}$ separately.

\subsection{Coupling to external electric field}\label{sec:Electric field coupling}

\subsubsection{Multipole expansion}

We will start our discussion with the energy shifts caused by the static electric field. In order to evaluate these shifts, let us first expand the electric potential

\begin{equation}
    \mathcal{H}_{el} = -\sum_{i=1}^N \phi(\bm{r}_i) = \sum_{k=0}^\infty \sum_{q=-k}^k I_{kq} \cdot \Theta_q^{(k)},
    \label{electric_interaction_hamiltonian}
\end{equation}

in terms of multipole operators

\begin{equation}
    \Theta_q^{(k)} = -\sum_{i=1}^N r_i^k C_{kq}(\theta_i, \varphi_i),
    \label{multipole_operators}
\end{equation}

that are constructed as irreducible tensors of rank $k$, see Ref. \cite{jackson}. In Eqs. (\ref{electric_interaction_hamiltonian})--(\ref{multipole_operators}), $\bm{r}_i = (r_i, \theta_i, \varphi_i)$ are the spherical coordinates of the i-th electron, $C_{kq} = \sqrt{4\pi / (2k+1)} Y_{kq}$ are the spherical harmonics in Racah normalization and

\begin{equation}
    I_{kq} = -\frac{1}{4\pi} \int_{\mathbb{R}^3} d^3\bm{r'}\; \frac{\triangle \phi(\bm{r'})}{r'^{k+1}}\, C_{kq}^*(\theta',\phi')
    \label{interior_multipole_moments}
\end{equation}

are the so-called interior multipole moments of the external charge distribution, which causes the electric potential $\phi$.

With the help of the multipole expansion (\ref{electric_interaction_hamiltonian})--(\ref{multipole_operators}) and by employing first-order perturbation theory one can evaluate the shifts of ionic levels caused by static external electric fields:

\begin{align}
\begin{split}
    \Delta E &\approx \langle \Gamma J M | \mathcal{H}_{el} | \Gamma J M \rangle \\
    &= \frac{1}{\sqrt{2J+1}} \sum_{k=0}^\infty \langle \Gamma J || \Theta_q^{(k)} || \Gamma J \rangle \sum_{q=-k}^k \langle J M\, k q | J M \rangle I_{kq},
    \label{ext_electric_energy_shift}
\end{split}
\end{align}

where the summation runs over all multipole components of $\phi$. Here we have used the Wigner-Eckart theorem following the convention of Ref. \cite{varshalovich} to express $\Delta E$ as a product of reduced matrix elements $\langle \Gamma J || \Theta_q^{(k)} || \Gamma J \rangle$ of multipole operators and the interior multipole moments $I_{kq}$. While the reduced matrix elements depend solely on the atomic structure, $I_{kq}$ reflects the behaviour of the external electric field at the position of the ion.

For the further evaluation of the energy shift $\Delta E$, it is convenient to rewrite Eq. (\ref{interior_multipole_moments}) in terms of the electric field and its derivatives at the position of the ion, placed at the coordinate origin. To achieve this goal, we can use Green's second identity for functions vanishing at infinity, $\langle \triangle \psi | \phi \rangle = \langle \psi | \triangle \phi \rangle$, and the relation

\begin{equation}
    \triangle(r^{-l-1} Y_{lm}) = \frac{(-1)^{l-1} 4\pi}{(2l-1)!!} Y_{lm}(\bm{\nabla})\delta,
    \label{Laplacian_relation}
\end{equation}

which follows from Lemma 3 of Ref. \cite{stampfer}. By using Eqs. (\ref{interior_multipole_moments})--(\ref{Laplacian_relation}) and performing some algebra, we find:

\begin{equation}
    I_{kq} = \frac{1}{(2l-1)!!} \left[ C_{lm}^*(\bm{\nabla}) \phi \right](0),
    \label{interior_multipole_moments_alt}
\end{equation}

where $C_{lm}^*(\bm{\nabla})$ in Eq. (\ref{interior_multipole_moments_alt}) or $Y_{lm}(\bm{\nabla})$ in Eq. (\ref{Laplacian_relation}) should be understood as plugging $\bm{\nabla} = (\partial_x, \partial_y, \partial_z)$ into the Cartesian representation of $r^l Y_{lm}(x, y, z)$, which is a polynomial in $x,y,z$. For instance $C_{1 -1}(\bm{\nabla}) = \frac{1}{\sqrt{2}} (\partial_x -i \partial_y)$, since $r^1C_{1 -1}(x, y, z) = \frac{1}{\sqrt{2}} (x - iy)$ and so on. The resulting expression is a differential operator, that acts in Eqs. (\ref{Laplacian_relation})--(\ref{interior_multipole_moments_alt}) on the delta distribution $\delta$ or the scalar potential $\phi$, respectively.

\subsubsection{Stark effect}

As we mentioned already above, the first-order energy shift (\ref{ext_electric_energy_shift}) of atomic levels can be represented as a sum over the different multipole components of the external field and electron distribution. In what follows, we will discuss the lowest-order terms of this expansion, which are known to yield relevant contributions to $\Delta E$. Since the monopole term, $k=0$, leads to a constant shift of all levels and hence, does not affect the transition energies, we start our discussion with the dipole term $k=1$.

For the dipole case, the interior multipole moment of the external charge distribution

\begin{equation}
    I_{1q} = \begin{cases}
        -\mathcal{E}_z(0) & q=0\\[5pt]
        \pm \frac{1}{\sqrt{2}} (\mathcal{E}_x(0) \mp i \mathcal{E}_y(0)) & q=\pm 1
    \end{cases},
\end{equation}

can be written in terms of spherical components of the field strength. Together with Eqs. (\ref{electric_interaction_hamiltonian})--(\ref{multipole_operators}), this leads to the well-known Stark shift operator

\begin{equation}
    \mathcal{H}_{el}^{k=1} = \mathcal{E}^{(1)} \cdot \Theta^{(1)} = \sum_{q=-1}^1 \mathcal{E}^{(1)}_q \Theta^{(1)}_{-q},
\end{equation}

where $\mathcal{E}^{(1)}$ denotes the electric field strength at the origin and $\Theta^{(1)}$ is the electric dipole operator. Both $\mathcal{E}^{(1)}$ and $\Theta^{(1)}$ are written here in the spherical basis. The theoretical treatment of this effect is extensively covered in Refs. \cite{angel_sandars, polarazibility, ba_polarazibilities} and for the sake of shortness will not be discussed here.

\subsubsection{Electric quadrupole shift}

In order to discuss the quadrupole contribution to the energy shift $\Delta E$, we start again with the interior multipole moment

\begin{equation}
    I_{2q} = \begin{cases}
        -\frac{1}{2}\partial_z \mathcal{E}_z(0) & q=0\\[5pt]
        \pm \frac{\sqrt{6}}{6} (\partial_x \mp i\partial_y)\mathcal{E}_z(0) & q=\pm 1 \\[5pt]
        -\frac{\sqrt{6}}{12} (\partial_x \mp i\partial_y)(\mathcal{E}_x \mp i\mathcal{E}_y)(0) & q = \pm 2
    \end{cases},
    \label{I_2q}
\end{equation}

which one can arrive at by evaluating Eq. (\ref{interior_multipole_moments_alt}) for $k=2$. By inserting Eq. (\ref{I_2q}) into the interaction Hamiltonian (\ref{electric_interaction_hamiltonian}), we obtain

\begin{equation}
    \mathcal{H}_{el}^{k=2} = \Theta^{(2)} \cdot \nabla\mathcal{E}^{(2)}, 
\end{equation}

where $\Theta^{(2)}$ is the atomic quadrupole operator and the spherical components of $\nabla \mathcal{E}^{(2)}$ can be written as

\begin{subequations}
\begin{align}
    \nabla \mathcal{E}^{(2)}_0 &= -\frac{1}{2}\partial_z \mathcal{E}_z(0), \\
    \nabla \mathcal{E}^{(2)}_{\pm 1} &= \pm \frac{\sqrt{6}}{6} (\partial_x \pm i\partial_y)\mathcal{E}_z(0), \\
    \nabla \mathcal{E}^{(2)}_{\pm 2} &= -\frac{\sqrt{6}}{12} (\partial_x \pm i\partial_y)(\mathcal{E}_x \pm i\mathcal{E}_y)(0).
    \label{spherical_tensor_field_gradient}
\end{align}
\end{subequations}

These components can be interpreted as the irreducible tensor representation of the electric field gradient and they readily show that the electric quadrupole shift can only occur in a non-homogeneous field.

Typically, the energy shift caused by the quadrupole interaction is very small, thus allowing one to treat the operator $\mathcal{H}_{el}^{k=2}$ within the first-order perturbation theory:

\begin{widetext}
\begin{align}
\begin{split}
    \Delta E_{EQS} &= \langle \Gamma J M | \mathcal{H}_{el}^{k=2} | \Gamma J M \rangle \\[10pt]
    &= \frac{\langle \Gamma J || \Theta^{(2)} || \Gamma J\rangle}{\sqrt{2J+1}} \langle J M\; 2 0 |J M\rangle \cdot A\left[ -3 \cos^2(\beta) + 1 + \epsilon\; \cos(2\alpha) \sin^2(\beta)\right].
    \label{EQS}
\end{split}
\end{align}
\end{widetext}

As seen from this formula, the quadrupole shift can be written as the product of the atomic part, that arises from the Wigner-Eckart theorem and of the term, that depends on the strength and geometry of the external field.
The atomic part is directly related to the quadrupole moment

\begin{equation}
    \Theta(\Gamma J) = \langle \Gamma J J | \Theta^{(2)}_0 | \Gamma J J \rangle
    \label{quadrupole_moment}
\end{equation}

of an atomic state $|\Gamma J \rangle$ by the trivial relation

\begin{equation}
    \langle \Gamma J || \Theta^{(2)} || \Gamma J\rangle = \frac{\sqrt{2J+1}}{\langle J J\, 2 0| J J \rangle} \Theta(\Gamma J).
\end{equation}

The evaluation of the quadrupole moment $\Theta(\Gamma J)$ can be done by employing the explicit formula of the electric quadrupole operator,

\begin{equation}
    \Theta_q^{(2)} = -\sum_{i=1}^N r_i^2 C_{2q}(\theta_i, \varphi_i),
    \label{quadrupole_operator}
\end{equation}

and using Racah algebra as shown in the appendix. Not much has to be discussed about the term, that describes the strength and geometry of the external field in Eq. (\ref{EQS}), since it is specific for each experimental setup. We refer the reader to Ref. \cite{itano} and just briefly note, that by rotating the quantisation axis to the principle axis frame of $\frac{\partial \phi}{\partial x_i \partial x_j}$ by the Euler angles $\alpha, \beta$, the quadrupole part of the electric field can be expressed as

\begin{equation}
    \phi^{(2)} = A\left[ (x'^2 + y'^2 - 2z'^2) + \epsilon (x'^2 - y'^2) \right],
    \label{quadrupole_potential}
\end{equation}

see Ref. \cite{quadrupole_pot}, where $A=U/r_0^2$ is the amplitude of the field gradient with a characteristic length $r_0$ and $\epsilon$ is a dimensionless asymmetry parameter.

\subsection{Coupling to external magnetic field}\label{sec:Magnetic field coupling}

\subsubsection{Homogeneous magnetic fields}

The coupling of an ion to an external magnetic field is described in Hamiltonian (\ref{Hamiltonian_external_fields}) by the operator

\begin{equation}
    \mathcal{H}_{mag} = \sum_{i=1}^N c \bm{\alpha_i} \cdot \bm{A}(\bm{r}_i, t),
    \label{magnetic_interaction_hamiltonian}
\end{equation}

where $\bm{A}$ is the magnetic vector potential. In what follows, we will further evaluate this interaction operator for the case of a static and homogeneous magnetic field as is appropriate for the analysis of "ion-in-trap" experiments. For this case, the vector potential can be expressed as

\begin{equation}
    \bm{A} = \frac{1}{2} (\bm{B} \times \bm{r}),
\end{equation}

which is referred to as the symmetric gauge \cite{symmetric_gauge}. By using this gauge and cyclic symmetry of the triple product, we can rewrite the interaction Hamiltonian as

\begin{equation}
    \mathcal{H}_{mag} = \frac{1}{2}\sum_{i=1}^N c\bm{B} \cdot (\bm{r}_i \times \bm{\alpha}_i).
\end{equation}

As a next step, we express the vector product $(\bm{r} \times \bm{\alpha})$ in terms of its irreducible tensor counterpart

\begin{equation}
    (\bm{r}\times \bm{\alpha})_q = -i\sqrt{2} r \left\{C^{(1)} \otimes \alpha^{(1)} \right\}^{(1)}_q,
\end{equation}

and arrive at the well-known expression

\begin{equation}
    \mathcal{H}_{mag} = \mu^{(1)} \cdot B^{(1)},
    \label{dipole_interaction_hamiltonian}
\end{equation}

where the magnetic dipole operator is given by

\begin{equation}
    \mu^{(1)}_q = -i c \frac{\sqrt{2}}{2} \sum_{i=1}^N r_i \left\{ C^{(1)}_i \otimes \alpha^{(1)}_i \right\}^{(1)}_q.
    \label{magnetic_dipole_operator}
\end{equation}

For a more accurate description of the interaction between an ion and magnetic field, one has to account also for the anomalous magnetic moment of the electron

\begin{equation}
    a = \frac{g_{\rm free}-2}{2},
\end{equation}

where $g_{\rm free} = 2.002 \ 319 \ 3...$ is the (free--electron) g--factor,
see Refs. \cite{anomalous_mag_moment, electron_g, anomalous_correction} for further details. This can be done by adding

\begin{equation}
    \Delta \mu^{(1)}_q = \mu_B \, \sum_{i=1}^N a \beta_i \Sigma^{(1)}_{i, q}
    \label{Delta_mu}
\end{equation}

to the magnetic dipole operator:

\begin{equation}
    \mathcal{H}_{mag} = (\mu^{(1)} + \Delta \mu^{(1)}) \cdot B^{(1)}.
    \label{dipole_interaction_hamiltonian2}
\end{equation}

Moreover, in Eq. (\ref{Delta_mu}), $\Sigma^{(1)}_{i, q}$ is the spherical tensor representation of the relativistic $4\times 4$ spin matrix

\begin{equation}
    \bm{\Sigma}_i = \begin{pmatrix}
        \bm{\sigma}_i & 0 \\
        0 & \bm{\sigma}_i
    \end{pmatrix}
\end{equation}

acting on the i-th electron.

\subsubsection{Linear Zeeman shift}

In order to estimate the energy shift caused by the magnetic interaction (\ref{dipole_interaction_hamiltonian2}), we again use a perturbative approach. To the first order of this theory and choosing the magnetic field along the quantisation axis, $\bm{B} = B\bm{e}_z$, we obtain:

\begin{equation}
    \Delta E_Z^{(1)} = g_J \mu_B M B,
\end{equation}

where the Landé g factor is

\begin{equation}
    g_J = 2\frac{\langle \Gamma J || \mu^{(1)} + \Delta \mu^{(1)} || \Gamma J\rangle }{\sqrt{J(J+1)(2J+1)}}.
\end{equation}

These expressions represent the well-known linear Zeeman shift, which is discussed in great detail in the literature \cite{lande-g-theory, astrophysics_lande_g}.

\subsubsection{Quadratic Zeeman shift}

In the following, we will treat the interaction Hamiltonian (\ref{dipole_interaction_hamiltonian2}) within the framework of second-order perturbation theory, to obtain the quadratic Zeeman shift of a reference state $|\Gamma J M \rangle$:

\begin{equation}
    \Delta E_Z^{(2)} = \sum_{\Gamma' J' M'} \frac{ | \langle \Gamma' J' M' | \mathcal{H}_{mag} | \Gamma J M \rangle |^2}{E(\Gamma J M) - E(\Gamma' J' M')}.
    \label{second_order_perturbation_theory}
\end{equation}

Here, the summation runs over all eigenstates $|\Gamma' J' M' \rangle$ of the unperturbed Hamiltonian (\ref{Dirac_Coulomb_Hamiltonian}) with energies $E(\Gamma' J' M')$ not equal to the energy $E(\Gamma J M)$ of the reference state.
We note, that for the unperturbed Hamiltonian (\ref{Dirac_Coulomb_Hamiltonian}), no splitting of the Zeeman sublevels occurs and hence, $E(\Gamma' J' M') = E(\Gamma' J')$ for all $M'$.
By inserting the operator (\ref{dipole_interaction_hamiltonian2}) into Eq. (\ref{second_order_perturbation_theory}) and making use of the Wigner-Eckart theorem, we obtain

\begin{equation}
    \Delta E_Z^{(2)} = C_2 \cdot B^2,
\end{equation}

where the quadratic Zeeman shift coefficient (QZSC) is given by

\begin{align}
\begin{split}
    C_2 = \sum_{\Gamma', J'} & \frac{1}{2J' + 1}\, \frac{|\langle J M\; 1 0 | J' M\rangle|^2}{E(\Gamma J) - E(\Gamma' J')} \\
    & \times |\langle \Gamma' J'||\mu^{(1)} + \Delta \mu^{(1)} ||\Gamma J\rangle|^2.
    \label{QZSC}
\end{split}
\end{align}

As seen from this expression, the summation over the entire atomic spectrum $|\Gamma' J' M' \rangle$ with $J' = J-1, J, J+1$ might be generally required for the accurate determination of $C_2$. However, for the calculations discussed in the present manuscript only few intermediate states with energies $E(\Gamma' J')$ close to the one of the reference state give a major contribution to the $C_2$ coefficient.

\section{\label{sec:numerical_methods}Numerical Methods}\label{sec:Numerical Methods}

\subsection{Atomic wavefunctions}

As seen from the discussion above, the calculations of both electric quadrupole and quadratic Zeeman shifts can be traced back to the evaluation of reduced matrix elements $\langle \Gamma J || T^{(k)} || \Gamma J \rangle$ with $T^{(2)} = \Theta^{(2)}$ and $T^{(1)} = \mu^{(1)} + \Delta \mu^{(1)}$, respectively. For many-electron systems, both operators can be represented as a sum $\sum_{i=1}^N t_i^{(k)}$ of one-electron operators, that are constructed as irreducible tensor operators of rank $k$.

For the evaluation of the reduced matrix element $\langle \Gamma J || T^{(k)} || \Gamma J \rangle$, we will employ the well-established multiconfiguration Dirac-Fock approach. Within this approach, the atomic state functions $|\Gamma J M \rangle$ can be written as a superposition

\begin{equation}
    |\Gamma J M \rangle = \sum_{\gamma} c_{\gamma} |\gamma J M \rangle
    \label{CI_expansion}
\end{equation}

of configuration state functions (CSFs) $| \gamma J M \rangle$, which are constructed as properly coupled Slater-determinants for particular electronic configurations $n_1\kappa_1, n_2\kappa_2, \dots, n_N\kappa_N$ and exhibit a total angular momentum $J$ and its projection $M$.
By employing a CSF expansion (\ref{CI_expansion}), the reduced matrix element can be written as

\begin{equation}
     \langle \Gamma J || T^{(k)} || \Gamma J\rangle = \sum_{\gamma, \gamma'} c_{\gamma} c_{\gamma'} \sum_{a,b} \zeta_{ab;k} \langle n_a \kappa_a || t^{(k)} || n_b \kappa_b\rangle,
     \label{MCDF_reduced_me}
\end{equation}

where $\zeta_{ab;k}$ are the so-called spin-angular coefficients \cite{grant, gaigalas}. Moreover, on the right-hand side of Eq. (\ref{MCDF_reduced_me}), $\langle n_a \kappa_a || t^{(k)} || n_b \kappa_b\rangle$ is the reduced matrix element of a one-electron operator, whose evaluation requires the use of angular momentum algebra and can be found in Appendix \ref{sec:one-electron-mes}.

Practical realization of Eq. (\ref{MCDF_reduced_me}) and hence, computation of the electric quadrupole moments and QZSCs requires the application of atomic structure packages. In the present study, we employ the GRASP \cite{grasp} and JAC \cite{JAC} programs, both of which are based on the relativistic (multiconfiguration) Dirac-Fock approach. For the latter code, we have developed and implemented the modules for the computation of quadrupole moments and quadratic Zeeman shift coefficients. An example for the usage of these modules, for instance \textit{LandeZeeman}, is presented in Appendix \ref{sec:JAC_example}.

\subsection{Computational details}

As seen from Eq. (\ref{CI_expansion}), an atomic state function is represented as an expansion in terms of configuration state functions, each of which is associated with a particular electronic configuration. This representation requires the determination of the radial parts of the CSFs as well as of the mixing coefficients $c_\gamma$.
For the computations below, the radial parts of the CSFs are optimized with respect to the Dirac-Coulomb Hamiltonian, where the electron-electron interaction $V_{ee}$ in Eq. (\ref{Dirac_Coulomb_Hamiltonian}) contains only the Coulomb term. The mixing coefficients $c_\gamma$ in Eq. (\ref{CI_expansion}) are obtained by diagonalizing the Hamiltonian matrix in the basis of CSFs. The latter allows for the proper account of electron-electron interactions beyond the Coulomb-Dirac-Fock approach, of the Breit interactions and self-energy QED corrections.

\begin{figure}[b]
    \centering
    \includegraphics[width=\linewidth]{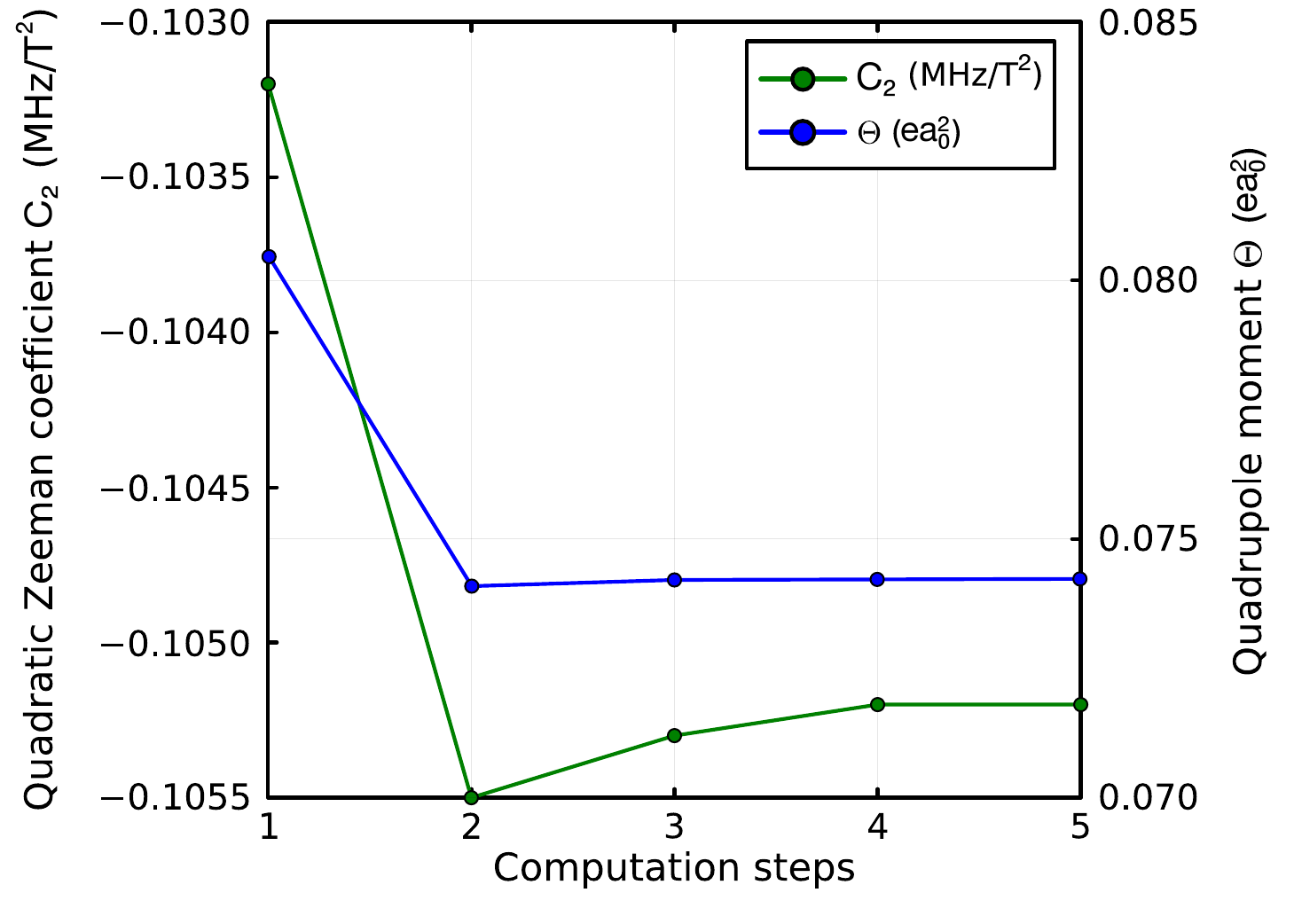}
    \caption{Electric quadrupole moment $\Theta$ (blue circles) and quadratic Zeeman shift coefficient $C_2$ (green circles) for the $^3P_2$ level of \ce{Ni^{12+}} ion. The $C_2$ coefficient is given with respect to the $M = 0$ substate. Results of the MCDF calculations are presented for different sizes of the active space of CSFs, as specified by the computational steps outlined in Table \ref{tab:Ni_Mg_calculation}.}
    \label{fig:Ni-convergence}
\end{figure}

For computational purposes, the expansion (\ref{CI_expansion}) requires the use of a suitable \textit{finite} basis set of configuration state functions, which is usually obtained by using a restricted active space (RAS) approach. Within the RAS, the CSF space is constructed using virtual excitations from one (or several) reference electronic configurations. 
In the present calculations, the single (S) and double (D) excitations of electrons from the valence shells have been taken into account (see Table \ref{tab:active_spaces}), which leads to a basis size of about $10^4 - 10^6$ CSFs. By analyzing the convergence of this expansion, we find that for most cases an accuracy of $\lesssim 1\, \%$ can be achieved for the quadrupole moment $\Theta$ and quadratic Zeeman shift coefficient $C_2$, which is sufficient for present applications. For example, in Fig. \ref{fig:Ni-convergence} we illustrate the convergence of our predictions for the $^3P_2$ level of \ce{Ni^{12+}} ion. The results are presented for different computational steps, as specified in Table \ref{tab:Ni_Mg_calculation}. As seen from the figure, a good convergence of $C_2$ and $\Theta$ coefficients can be observed at the fourth computational step, which refers to single and double excitations from the reference configuration [Ne]~3s$^2$~3p$^4$ into orbitals with principle quantum numbers up to $n=6$. Similar or better convergence was observed also for other systems considered in the present work.
Furthermore, we have also performed test calculations to treat virtual excitations of the core electrons perturbatively. These computations revealed that the core-core and core-valence contributions are small and can be neglected within the context of the present study.

\begin{table}
    \caption{Reference configurations and active spaces as applied in MCDF calculations of various HCIs. In all computations, single and double virtual excitations from the valence orbitals were considered. %The last column of the table shows the highest principle quantum number, to which virtual excitations were allowed. 
    Moreover, all computations included Breit and QED corrections using a configuration interaction approach.}
    \begin{ruledtabular}
    \begin{tabular}{c|c|c}
        Ion & Reference configuration & Active space \\
        \hline
        \ce{Ca^{14+}} & $[\ce{He}] 2s^2 2p^2$ & SD excitations up to $5spdfg$ \\
        \ce{Ni^{12+}} & $[\ce{Ne}] 3s^2 3p^4$ & SD excitations up to $7spdfg$ \\
        \ce{Xe^{9+}} & $[\ce{Kr}] 4d^9$ & SD excitations up to $7spdfg$ \\
        \ce{Xe^{10+}} & $[\ce{Kr}] 4d^8$ & SD excitations up to $7spdfg$ \\
        \ce{Xe^{11+}} & $[\ce{Kr}] 4d^7$ & SD excitations up to $7spdfg$ \\
        \ce{Xe^{12+}} & $[\ce{Kr}] 4d^6$ & SD excitations up to $7spdfg$ \\
        \ce{Xe^{15+}} & $[\ce{Kr}] 4d^3$ & SD excitations up to $7spdfg$ \\
        \ce{Xe^{16+}} & $[\ce{Kr}] 4d^2$ & SD excitations up to $7spdfg$ \\
        \ce{Xe^{17+}} & $[\ce{Kr}] 4d^1$ & SD excitations up to $7spdfg$ \\
    \end{tabular}
    \label{tab:active_spaces}
    \end{ruledtabular}
\end{table}

\begin{table*}
    \caption{Second order Zeeman shift coefficients and electric quadrupole moments for $^3P_0$ level of neutral \ce{Mg} atom as well as for $^3P_2$ level of \ce{Ni^{12+}} ion. The $C_2$ coefficients were computed for the $M = 0$ substates. Results of our MCDF calculations are presented for various computational steps, which correspond to different active spaces of CSFs as specified in the second column.}
    \begin{ruledtabular}
    \begin{tabular}{ccccc}
        Computational step & Active space & \multicolumn{2}{c}{$C_2$ (MHz/T$^2$)} & $\Theta$ ($ea_0^2$) \\
        \cline{3-5}
         & & \ce{Mg} & \ce{Ni^{12+}} & \ce{Ni^{12+}} \\
        \hline
        1 & Dirac Fock, no virtual excitations & -206.9 & -0.1032 & 0.08046 \\
        2 & SD excitations from $\{3s,3p\}$ up to $4spdf$ & -228.3 & -0.1055 & 0.07409 \\
        3 & SD excitations from $\{3s,3p\}$ up to $5spdfg$ & -228.3 & -0.1053 & 0.07421 \\
        4 & SD excitations from $\{3s,3p\}$ up to $6spdfg$ & -228.6 & -0.1052 & 0.07422 \\
        5 & SD excitations from $\{3s,3p\}$ up to $7spdfg$ & -228.5 & -0.1052 & 0.07423 \\
    \end{tabular}
    \label{tab:Ni_Mg_calculation}
    \end{ruledtabular}
\end{table*}

\section{\label{sec:results}Results and Discussion}\label{sec:Results}

The theory discussed in the previous sections can be applied to arbitrary many-electron systems, assuming a proper representation of the ASFs (\ref{CI_expansion}).
In Tables \ref{tab:qzsc_results} and \ref{tab:qm_results} for example, we revisited previous experimental and theoretical data for the QZSCs and quadrupole moments of several neutral atoms and singly charged ions. We performed calculations of these systems, mainly to probe the accuracy of our numerical approach.
In Table \ref{tab:qzsc_results}, we present the QZSCs calculated for $^3P_0$ states of \ce{Al^+}, \ce{Mg} and \ce{Sr} atoms. In these calculations, the wave functions of both the reference $^3P_0$ and intermediate $^3P_1$ and $^1P_1$ states were generated within the MCDF approach as implemented in the GRASP code. Moreover, as seen from Eq. (\ref{QZSC}), the evaluation of the QZSC also requires knowledge of the energies $E(\Gamma J)$ and $E(\Gamma' J')$. These energies were obtained either from our MCDF calculations or taken from the NIST atomic spectra database \cite{NIST_ASD}. As seen from Table \ref{tab:qzsc_results}, the latter approach provides a better agreement with previous theoretical results. Our calculations also agree quite well with the experimental data, except for \ce{Mg}, where all computations including previous ones overestimate $C_2$.

\begin{table*}
    \caption{The quadratic Zeeman shift coefficients of $^3P_0$ states of \ce{Al^+}, \ce{Mg} and \ce{Sr} atoms. The results of our MCDF calculations based on Eq. (\ref{QZSC}) are presented in the third and fourth columns. The energies, that enter in the denominator of Eq. (\ref{QZSC}) have been taken from our MCDF calculations and from the NIST atomic spectra database \cite{NIST_ASD}, respectively. Moreover, our results are compared with previous experimental and theoretical data shown in the last column.}
    \begin{ruledtabular}
    \begin{tabular}{lccclr}
         & & \multicolumn{4}{c}{QZSC $C_2$ (MHz/T$^2$)} \\
        \cline{3-6}
        Element & State & MCDF & MCDF + NIST ASD & \multicolumn{2}{c}{Previous Data} \\
        \hline
        \multirow{2}{*}{\ce{^{27}Al^{+}}} & \multirow{2}{*}{$[\ce{Ne}]3s3p\;^3P_0$} & \multirow{2}{*}{-74.2(2.2)} & \multirow{2}{*}{-71.9(2.2)} & Experiment \cite{al+_qzsc}: & -71.944(24)\\
        & & & & Theory \cite{al+_qzsc}: & -71.927 \\
        \hline
        \multirow{3}{*}{\ce{^{25} Mg}} & \multirow{3}{*}{$[\ce{Ne}]3s3p\;^3P_0$} & \multirow{3}{*}{-228(6)} & \multirow{3}{*}{-218(6)} & Experiment: & $-206.6(2.0)$ \\
         & & & & Theory \cite{Mg_Taichenachev}: & $-217$ \\
         & & & & Theory \cite{Mg_QZSC_chin}: & $-216(1)$ \\
        \hline
        \multirow{3}{*}{\ce{^{87} Sr}} & \multirow{3}{*}{$[\ce{Kr}]5s5p\;^3P_0$} & \multirow{3}{*}{-25.9(7)} & \multirow{3}{*}{-23.408(4)} & Experiment \cite{Sr_6}: & $-23.38(3)$ \\
         & & & & Theory \cite{Mg_Taichenachev}: & $-23.3$ \\
         & & & & Theory \cite{Sr_theory}: & $-23.38(5)$ \\
    \end{tabular}
    \end{ruledtabular}
    \label{tab:qzsc_results}
\end{table*}

Apart from the QZSCs, we have also performed a number of calculations for the electric quadrupole moments of singly charged \ce{Sr^+}, \ce{Yb^+} and \ce{Hg^+} ions. The results of these calculations are presented in Table \ref{tab:qm_results}. Again, we observe a good agreement with previous theoretical and experimental data, except for the case of the $^2F_{7/2}$ state of \ce{Yb^+}. For this case, all theoretical calculations overestimate the absolute value of the quadrupole moment, which can be attributed to a very complex electronic structure of \ce{Yb^+}.
We gather from the comparison with previous results, that even for rather complicated systems, our computations seem to provide reliable outcomes. This gives us confidence in our results below for highly charged ions, which typically have a simpler electronic structure.

\begin{table*}
    \caption{The quadrupole moment of the \ce{^{88}Sr^+} $^2D_{5/2}$, \ce{^{171}Yb^+} $^2D_{3/2}$, \ce{^{171}Yb^+} $^2F_{7/2}$ and \ce{^{199}Hg^+} $^2D_{5/2}$ states. The results based on our MCDF calculations are presented in the third column. Moreover, our results are compared with previous theoretical and experimental data shown in the last column.}
    \begin{ruledtabular}
    \begin{tabular}{lcclr}
         & & \multicolumn{3}{c}{Quadrupole Moment $\Theta$ ($e a_B^2$)} \\
        \cline{3-5}
        Element & Level & MCDF & \multicolumn{2}{c}{Previous Data} \\
        \hline
        \multirow{2}{*}{\ce{^{88}Sr^+}} &  \multirow{2}{*}{$[\ce{Kr}]4d\; ^2D_{5/2}$} &  \multirow{2}{*}{3.03(12)} & Experiment \cite{Sr+_EQS}: & 2.6(3) \\
         & & & Theory \cite{Sr+_EQS}: & 3.0 \\
        \hline
        \multirow{2}{*}{\ce{^{171}Yb^+}} & \multirow{2}{*}{$[\ce{Xe}] 4f^{14} 5d\; ^2D_{3/2}$} & \multirow{2}{*}{1.9(4)} & Experiment \cite{Yb+_2D3/2}: & 2.08(11) \\
         & & & Theory \cite{Yb+_theory}: & 2.068(12) \\
        \hline
        \multirow{2}{*}{\ce{^{171}Yb^+}} & \multirow{2}{*}{$[\ce{Xe}] 4f^{13} 6s^2\; ^2F_{7/2}$} & \multirow{2}{*}{-0.16(3)} & Experiment \cite{Yb+_2F7/2}: & -0.041(5) \\
         & & & Theory \cite{Yb+_theory}: & -0.216(20) \\
        \hline
        \multirow{2}{*}{\ce{^{199}Hg^+}} & \multirow{2}{*}{$[\ce{Xe}] 4f^{14} 5d^9 6s^2\; ^2D_{5/2}$} & \multirow{2}{*}{-0.55(8)} & Experiment \cite{Hg+_quadrupole_moment}: & -0.510(18) \\
        & & & Theory \cite{Hg+_quadrupole_moment}: & -0.5440 \\
    \end{tabular}
    \end{ruledtabular}
    \label{tab:qm_results}
\end{table*}

In Table \ref{tab:ca_qzsc_qm}, we present our calculations of QZSCs and quadrupole moments for all Zeeman sublevels of $^3P_{0,1,2}$ states of \ce{Ca^{14+}} and \ce{Ni^{12+}} ions.
These ions are of special interest for high-precision spectroscopy experiments and as candidates for HCI optical clocks, respectively, and their response to external electric and magnetic fields has to be well understood.
As seen from the table, the predicted values for both $\Theta$ and $C_2$ coefficients are about two orders of magnitude smaller than the values for neutral atoms or singly charged ions, presented in Tables \ref{tab:qzsc_results} and \ref{tab:qm_results}.
These results confirm our expectation of a reduced sensitivity of highly charged ions to external perturbations \cite{HCI_transitions}, which makes them a convenient tool for fundamental physics research and optical clocks.
For example, an accuracy of about $0.1\, \si{\Hz}$ has been achieved recently in measurements of the $|^3P_2,\, M=0\rangle \rightarrow |^3P_0,\, M=0\rangle$ absolute frequency in \ce{Ni^{12+}} ions \cite{Ni_absolute_uncertainty}.
For this transition, we estimate the electric quadrupole and second-order Zeeman shifts to be $\Delta E_{EQS} \approx -67(19) \, \si{\milli\Hz}$ and $\Delta E_{Z}^{(2)} \approx 3.1(8)\, \si{\milli\Hz}$, respectively. These estimates are based on using typical parameters \cite{Ar13+} of the rf ion traps, $A= 2.0\, \si{\V\per\square\milli\m}$, $\epsilon = 1$, $\alpha = 45(5)^{\circ}$, $\beta = 30(5)^{\circ}$, $B=20\, \si{\micro\tesla}$, see Eq. (\ref{quadrupole_potential}).
\begin{table*}
    \caption{The quadrupole moments and quadratic Zeeman shift coefficients of $^3P_{0,1,2}$ states of \ce{Ca^{14+}} and \ce{Ni^{12+}} ions. The results of our MCDF calculations based on Eqs. (\ref{quadrupole_moment}) and (\ref{QZSC}) are presented in the third, fifth and sixth columns. The energies, that enter in the denominator of Eq. (\ref{QZSC}) have been taken from our MCDF calculations and from the NIST atomic spectra database \cite{NIST_ASD}, respectively.}
    \centering
    \begin{ruledtabular}
    \begin{tabular}{cccccc}
         & & & & \multicolumn{2}{c}{QZSC $C_2$ (MHz/T$^2$)} \\
        \cline{5-6}
        Ion & State & Quadrupole Moment $\Theta$ ($a_0^2e$) & $M$ & MCDF & MCDF + NIST ASD \\
        \hline
        \multirow{6}{*}{\ce{Ca^{14+}}} & $[\ce{He}] 2s^22p^2\;^3P_0$ & 0 & 0 & -0.2447(14) & -0.2435(2) \\
        \cline{2-6}
         & \multirow{2}{*}{$[\ce{He}] 2s^22p^2\;^3P_1$} & \multirow{2}{*}{0.01018(1)} & $\pm 1$ & -0.0761(4) & -0.0851(1) \\
         & & & $0$ & 0.143(2) & 0.1295(3) \\
        \cline{2-6}
         & \multirow{3}{*}{$[\ce{He}] 2s^22p^2\;^3P_2$} & \multirow{3}{*}{-0.0165(1)} & $\pm 2$ & -0.0049(1) & -0.0050(1) \\
         & & & $\pm 1$ & 0.0738(4) & 0.0828(1) \\
         & & & 0 & 0.1001(6) & 0.1121(1) \\
        \hline
        \multirow{6}{*}{\ce{Ni^{12+}}} & \multirow{3}{*}{$[\ce{Ne}] 3s^2 3p^4\;^3P_2$} & \multirow{3}{*}{0.074(3)} & $\pm 2$ & -0.0084(5) & -0.0084(5) \\
         & & & $\pm 1$ & -0.0802(7) & -0.0806(7) \\
         & & & $0$ & -0.1052(9) & -0.1046(9) \\
        \cline{2-6}
         & \multirow{2}{*}{$[\ce{Ne}] 3s^2 3p^4\;^3P_1$} & \multirow{2}{*}{-0.046(1)} & $\pm 1$ & 0.0743(7) & 0.0746(7) \\
         & & & $0$ & -5.9(1.8) & -7.5(1.8) \\
         \cline{2-6}
         & $[\ce{Ne}] 3s^2 3p^4\;^3P_0$ & 0 & 0 & 6.0(1.8) & 7.6(1.8) \\
    \end{tabular}
    \end{ruledtabular}
    \label{tab:ca_qzsc_qm}
\end{table*}

In addition to \ce{Ca^{14+}} and \ce{Ni^{12+}}, highly charged \ce{Xe} ions also attract a particular interest for isotope shift studies due to the abundance of optical transitions and the highest number of stable even isotopes among all elements \cite{Highly_charged_Xe}.
The electric quadrupole moments and QZSCs for the ground and some excited states of these ions are displayed in Table \ref{tab:xe_qzsc_qm_1}.
For the computation of $C_2$ coefficients, all fine-structure states belonging to the reference electronic configuration were used for intermediate state summation in Eq. (\ref{QZSC}). In the special cases of \ce{Xe^{9+}} and \ce{Xe^{17+}}, there are only two states, $^2D_{3/2}$ and $^2D_{5/2}$, that come with each of the $[\ce{Kr}] 4d^9$ and $[\ce{Kr}] 4d$ configurations. This fact together with the symmetry properties of the Clebsch-Gordan coefficient in Eq. (\ref{QZSC}) implies vanishing $C_2$ for the $^2D_{5/2}$ $M=\pm 5/2$ Zeeman sublevels. The QZSCs for all other sublevels lie in a range from 0.004 to 0.7 MHz/T$^2$, which, similar to \ce{Ca^{14+}} and \ce{Ni^{12+}} predictions, is much smaller than typical values for neutral atoms.

To make an additional test of our calculations, the MCDF predictions for $C_2$ and $\Theta$ coefficients from Table \ref{tab:xe_qzsc_qm_1} were compared with Dirac-Fock results obtained using the JAC package. A good agreement between DF and MCDF computations was identified for all $C_2$ and $\Theta$ coefficients.

\begin{table*}
    \centering
    \caption{The quadrupole moments and quadratic Zeeman shift coefficients of several highly charged \ce{Xe} ions. The results of our MCDF calculations based on Eqs. (\ref{quadrupole_moment}) and (\ref{QZSC}) are presented in the third and fifth columns. The energies, that enter in the denominator of Eq. (\ref{QZSC}) have been taken from our MCDF calculations.}
    \begin{ruledtabular}
    \begin{tabular}{ccccc}
         & & & \multicolumn{2}{c}{QZSC} \\
        \cline{4-5}
        Ion & State & Quadrupole Moment $\Theta$ ($a_0^2e$) & $M$ & $C_2$ (MHz/T$^2$) \\
        \hline
        \multirow{5}{*}{\ce{Xe^{9+}}} & \multirow{3}{*}{$[\ce{Kr}] 4d^9\; ^2D_{5/2}$} & \multirow{3}{*}{-0.205(9)} & $\pm 5/2$  & 0 \\
         & & & $\pm 3/2$ & -0.063(1) \\
         & & & $\pm 1/2$  & -0.095(2) \\
        \cline{2-5}
         & \multirow{2}{*}{$[\ce{Kr}] 4d^9\; ^2D_{3/2}$} & \multirow{2}{*}{-0.141(6)} & $\pm 3/2$ & 0.063(1) \\
         & & & $\pm 1/2$& 0.095(2) \\
        \hline
        \multirow{15}{*}{\ce{Xe^{10+}}} & \multirow{5}{*}{$[\ce{Kr}] 4d^8\; ^3F_4$} & \multirow{5}{*}{-0.115(3)} & $\pm 4$ & -0.0041(3) \\
          & & & $\pm 3$ & -0.0810(8) \\
          & & & $\pm 2$ & -0.136(2) \\
          & & & $\pm 1$ & -0.169(2) \\
          & & & $0$ & -0.180(2) \\
        \cline{2-5}
         & \multirow{3}{*}{$[\ce{Kr}] 4d^8\; ^3F_2$}  & \multirow{3}{*}{0.063(4)} & $\pm 2$ & -0.55(28) \\
         & & & $\pm 1$ & -0.7(5) \\
          & & & $0$ & -0.7(6) \\
        \cline{2-5}
         & \multirow{4}{*}{$[\ce{Kr}] 4d^8\; ^3F_3$} & \multirow{4}{*}{-0.080(2)} & $\pm 3$ & 0.0775(8) \\
          & & & $\pm 2$ & 0.4(3) \\
          & & & $\pm 1$ & 0.6(5) \\
          & & & $0$ & 0.7(5) \\
        \cline{2-5}
         & \multirow{5}{*}{$[\ce{Kr}] 4d^8\; ^1G_4$} & \multirow{5}{*}{-0.40(2)} & $\pm 4$ & 0.0041(2) \\
          & & & $\pm 3$ & 0.0036(2) \\
          & & & $\pm 2$ & 0.0032(2) \\
          & & & $\pm 1$ & 0.0030(2) \\
          & & & $0$ & 0.0029(2) \\
        \hline
        \multirow{14}{*}{\ce{Xe^{11+}}} & \multirow{5}{*}{$[\ce{Kr}] 4d^7\; ^4F_{9/2}$} & \multirow{5}{*}{0.079(3)} & $\pm 9/2$ & -0.0212(6)  \\
         & & & $\pm 7/2$ & -0.114(2) \\
         & & & $\pm 5/2$ & -0.184(2) \\
         & & & $\pm 3/2$ & -0.230(3) \\
         & & & $\pm 1/2$ & -0.253(3) \\
        \cline{2-5}
         & \multirow{4}{*}{$[\ce{Kr}] 4d^7\; ^4F_{7/2}$} & \multirow{4}{*}{0.060(2)} & $\pm 7/2$ & 0.090(2) \\
         & & & $\pm 5/2$ & -0.38(3) \\
         & & & $\pm 3/2$ & -0.70(4) \\
         & & & $\pm 1/2$ & -0.86(5) \\
        \cline{2-5}
         & \multirow{5}{*}{$[\ce{Kr}] 4d^7\; ^2G_{9/2}$} & \multirow{5}{*}{-0.091(2)} & $\pm 9/2$ & -0.033(1) \\
         & & & $\pm 7/2$ & -0.0723(8) \\
         & & & $\pm 5/2$ & -0.102(2) \\
         & & & $\pm 3/2$ & -0.121(2) \\
         & & & $\pm 1/2$ & -0.131(2) \\
        \hline
        \multirow{10}{*}{\ce{Xe^{12+}}} & \multirow{5}{*}{$[\ce{Kr}] 4d^6\; ^5D_4$} & \multirow{5}{*}{0.186(5)} & $\pm 4$ & -0.028(2) \\
         & & & $\pm 3$ & -0.170(2) \\
         & & & $\pm 2$ & -0.272(3) \\
         & & & $\pm 1$ & -0.334(4) \\
         & & & $0$ & -0.354(5) \\
        \cline{2-5}
         & \multirow{5}{*}{$[\ce{Kr}] 4d^6\; ^3H_4$} & \multirow{5}{*}{0.103(2)} & $\pm 4$ & -0.23(2) \\
         & & & $\pm 3$ & -0.28(2) \\
         & & & $\pm 2$ & -0.31(2) \\
         & & & $\pm 1$ & -0.33(2) \\
         & & & $0$ & -0.34(2) \\
    \end{tabular}
    \end{ruledtabular}
    \label{tab:xe_qzsc_qm_1}
\end{table*}

\begin{table*}
    \ContinuedFloat
    \caption{(continued)}
    \centering
    \begin{ruledtabular}
    \begin{tabular}{ccccc}
         & & & \multicolumn{2}{c}{QZSC} \\
        \cline{4-5}
        Ion & State & Quadrupole Moment $\Theta$ ($a_0^2e$) & $M$ & $C_2$ (MHz/T$^2$) \\
        \hline
        \multirow{5}{*}{\ce{Xe^{15+}}} & \multirow{2}{*}{$[\ce{Kr}] 4d^3\; ^4F_{3/2}$} & \multirow{2}{*}{-0.065(2)} & $\pm 3/2$ & -0.341(7) \\
         & & & $\pm 1/2$ & -0.49(2) \\
        \cline{2-5}
         & \multirow{2}{*}{$[\ce{Kr}] 4d^3\; ^4P_{3/2}$} & \multirow{2}{*}{-0.077(3)} & $\pm 3/2$ & -0.102(3) \\
         & & & $\pm 1/2$ & -0.59(3) \\
        \cline{2-5}
         & $[\ce{Kr}] 4d^3\; ^4P_{1/2}$ & 0 & $\pm 1/2$ & 0.43(3) \\
        \hline
        \multirow{12}{*}{\ce{Xe^{16+}}} & \multirow{3}{*}{$[\ce{Kr}] 4d^2\; ^3F_2$} & \multirow{3}{*}{0.039(2)} & $\pm 2$ & -0.114(2) \\
         & & & $\pm 1$ & -0.170(3) \\
         & & & $0$ & -0.188(3) \\
        \cline{2-5}
         &  \multirow{4}{*}{$[\ce{Kr}] 4d^2\; ^3F_3$} & \multirow{4}{*}{0.0667(7)} & $\pm 3$ & -0.079(1) \\
         & & & $\pm 2$ & -0.042(2) \\
         & & & $\pm 1$ & -0.020(3) \\
         & & & $0$ & -0.012(4) \\
        \cline{2-5}
         & \multirow{3}{*}{$[\ce{Kr}] 4d^2\; ^1D_2$} & \multirow{3}{*}{-0.0717(7)} & $\pm 2$ & -0.054(2) \\
         & & & $\pm 1$ & -0.143(8) \\
         & & & $0$ & -0.17(1) \\
        \cline{2-5}
         & \multirow{2}{*}{$[\ce{Kr}] 4d^2\; ^3P_1$} & \multirow{2}{*}{0.0630(1)} & $\pm 1$ & 0.094(8) \\
         & & & $0$ & 0.49(2) \\
        \hline
        \multirow{5}{*}{\ce{Xe^{17+}}} & \multirow{2}{*}{$[\ce{Kr}] 4d\; ^2D_{3/2}$} & \multirow{2}{*}{0.109(4)} & $\pm 3/2$ & -0.046(1) \\
         & & & $\pm 1/2$ & -0.069(1) \\
        \cline{2-5}
         & \multirow{3}{*}{$[\ce{Kr}] 4d\; ^2D_{5/2}$} & \multirow{3}{*}{0.159(6)} & $\pm 5/2$ & 0 \\
         & & & $\pm 3/2$ & 0.046(1) \\
         & & & $\pm 1/2$ & 0.069(1) \\
    \end{tabular}
    \end{ruledtabular}
    \label{tab:xe_qzsc_qm_2}
\end{table*}

\section{\label{sec:summary}Summary and Outlook}\label{sec:Summary}

The knowledge of electric quadrupole and quadratic Zeeman shifts is becoming increasingly crucial for high-precision spectroscopy. Of special interest here are measurements with highly charged ions, where these shifts are expected to be small \cite{HCI_transitions} and accurate calculations are possible.
In this contribution, therefore, we revisited the theory of the electric quadrupole and second-order Zeeman shifts based on the first and second-order perturbative approach. In particular, starting from the analysis of the Hamiltonian describing an atom exposed to external static electric and magnetic fields, we derived the general expressions for the $\Theta$ and $C_2$ coefficients, which are employed to parameterize both shifts. The computation of these coefficients within the framework of multiconfiguration Dirac-Fock (MCDF) was also discussed in detail.

To implement the MCDF approach for the calculation of quadrupole moments and QZSCs, we employed the well-established GRASP computer package as well as the newly emerging JAC toolbox. For the latter, we have developed the modules for the calculation of both, $\Theta$ and $C_2$, coefficients. By using the GRASP and JAC packages, detailed calculations have been performed for \ce{Ca^{14+}}, \ce{Ni^{12+}}, \ce{Xe^{q+}}, $q=9-17$, ions. These ions are currently of special interest for high-precision spectroscopic measurements, that aim, in particular, for the search of new physics beyond the standard model \cite{HCI, isotope_shift, Highly_charged_Xe}.
Our calculations contribute important results to the analysis and planning of such measurements carried out and planned at various sites.

The theory laid out in the present work is restricted to the analysis of quadrupole moments and QZSCs of fine-structure ionic levels, i.e. assuming zero nuclear spin. Currently, however, spectroscopy of hyperfine levels also attracts particular attention \cite{cf_clocks, pr_clock, fine_structure_variation, Bekker2019}. In what follows, we plan to extend the theory and program implementation to investigate shifts of hyperfine levels, which are presently not yet well-explored from a theoretical standpoint. This work is currently underway and the results will be published in forthcoming publications.

\appendix

\section{\label{sec:one-electron-mes}One electron reduced matrix elements}

The computation of QZSCs (\ref{QZSC}) and quadrupole moments (\ref{quadrupole_moment}) requires the evaluation of reduced matrix elements of the magnetic dipole operator (\ref{magnetic_dipole_operator}) and the electric multipole operator (\ref{quadrupole_operator}). Employing the MCDF approach, a many-electron reduced matrix element can be expressed in terms of its single-electron counterparts as shown in Eq. (\ref{MCDF_reduced_me}). All required one-electron matrix elements can be separated into a radial and an angular part, where the latter can be analytically determined using angular momentum algebra. This immensely reduces the computational load. 
In order to discuss such an analysis, we have to briefly recall the representation of one-electron solutions of the Dirac equation in a central potential. For bound electron states, the wavefunctions can be written as:

\begin{equation}
    \psi_{n \kappa m}(\bm{r}) = \frac{1}{r} \begin{pmatrix}
        P_{n\kappa}(r) \Omega_{\kappa m}(\theta, \varphi) \\
        i Q_{n\kappa}(r) \Omega_{-\kappa m}(\theta, \varphi)
    \end{pmatrix},
    \label{One-electron-bound-state}
\end{equation}

where $P_{n\kappa}(r)$ and $Q_{n\kappa}(r)$ are the large and small radial components and

\begin{equation}
    \Omega_{\kappa m}(\theta, \varphi) = \sum_{m_l,m_s} \langle l\, m_l\; 1/2\, m_s | j m \rangle Y_{lm}(\theta, \varphi) \chi_{m_s}
\end{equation}

are the generalized spherical harmonics with $\chi_{+1/2} = (1\: 0)^T$, $\chi_{-1/2} = (0\: 1)^T$ being standard two-component Pauli spinors.
The wave function (\ref{One-electron-bound-state}) is characterized by the principle quantum number $n$, Dirac quantum number

\begin{equation}
    \kappa := \begin{cases}
        j+\frac{1}{2} & \text{if } l = j + \frac{1}{2} \\
        -(j + \frac{1}{2}) & \text{if } l = j - \frac{1}{2}
    \end{cases},
\end{equation}

as well as orbital and total angular momenta $l$ and $j$, and the magnetic quantum number $m = -j, ..., j$.
By using the radial-angular representation (\ref{One-electron-bound-state}) of an electron wave function, we are ready to evaluate the single-electron reduced matrix elements, needed for the computations of $\Theta$ and $C_2$.

The one-electron reduced matrix element of the electric quadrupole operator $\Theta^{(2)}$ can be evaluated as:

\begin{equation}
    \langle n_a \kappa_a  || \Theta^{(2)} || n_b \kappa_b \rangle = -\langle r^2\rangle_{n_a\kappa_a, n_b\kappa_b} \cdot \langle \kappa_a || C^{(2)} ||\kappa_b \rangle,
    \label{quadrupole_single_electron_reduced_me}
\end{equation}

where the radial integral is given by

\begin{equation}
    \langle r^k\rangle_{n_a\kappa_a, n_b\kappa_b} = \int_0^\infty r^k (P_{n_a\kappa_a} P_{n_b \kappa_b} + Q_{n_a\kappa_a} Q_{n_b \kappa_b})\; dr,
    \label{k-order_radial_integral}
\end{equation}

and the angular part 

\begin{align}
\begin{split}
    &\langle \kappa_a || C^{(k)} || \kappa_b \rangle \\
    &= (-1)^k \sqrt{2j_a + 1}\, \langle j_a\, 1/2\; k\, 0 | j_b\, 1/2 \rangle\, \frac{1 + (-1)^{l_a + l_b + k}}{2}
    \label{reduced_C_k_me}
\end{split}
\end{align}

can be expressed in terms of a Clebsch-Gordan coefficient and a parity selection rule.

In contrast to the $\Theta$ coefficient, the QZSC requires the evaluation of reduced matrix elements of two operators, the magnetic dipole operator (\ref{magnetic_dipole_operator}) as well as the QED correction (\ref{Delta_mu}).
The reduced matrix elements of $\mu^{(1)}$ can be expressed as

\begin{align}
\begin{split}
    &\langle n_a \kappa_a || \mu^{(1)} || n_b \kappa_b \rangle \\
    &= - \frac{c}{2} [r]_{n_a \kappa_a n_b \kappa_b} (\kappa_a + \kappa_b) \langle -\kappa_a || C^{(1)} || \kappa_b \rangle,
    \label{mu_one_electron_reduced_me}
\end{split}
\end{align}

where the radial integral is

\begin{equation}
    [r^k]_{n_a \kappa_a n_b \kappa_b} = \int_0^\infty r^k (P_{n_a \kappa_a} Q_{n_b \kappa_b} + P_{n_b \kappa_b} Q_{n_a \kappa_a})\; dr.
\end{equation}

The reduced matrix elements of the QED correction to the magnetic dipole operator can be obtained as follows:

\begin{align}
\begin{split}
    &\langle n_a \kappa_a || \Delta \mu^{(1)} || n_b \kappa_b \rangle \\
    &= \frac{g-2}{2} \mu_B \bigg\{ \int_0^\infty P_{n_a \kappa_a} P_{n_b \kappa_b} dr\; \langle \kappa_a || \sigma^{(1)} || \kappa_b \rangle \\
    &\qquad - \int_0^\infty Q_{n_a \kappa_a} Q_{n_b \kappa_b} dr\; \langle -\kappa_a || \sigma^{(1)} || -\kappa_b \rangle \bigg\},
    \label{delta_mu_one_electron_reduced_me}
\end{split}
\end{align}

where

\begin{align}
\begin{split}
    \langle \kappa_a || \sigma^{(1)} || \kappa_b \rangle &= (-1)^{l_b + j_a + 3/2} \sqrt{6 (2j_b + 1) (2j_a + 1)} \\
    &\qquad \times \begin{Bmatrix}
        1/2 & l_b & j_b \\
        j_a & 1 & 1/2
    \end{Bmatrix}\, \delta_{l_a l_b}.
    \label{reduced_pauli_me}
\end{split}
\end{align}

\section{\label{sec:JAC_example}Computation of QZSCs and electric quadrupole moments using JAC}

The evaluation of QZSCs and electric quadrupole moments requires the use of a suitable approximation of atomic state functions. This task is tackled by several -- more or less -- user-friendly atomic structure packages.
To support the computation of second-order Zeeman and electric quadrupole coefficients $C_2$ and $\Theta$ we have implemented \textit{LandeZeeman}, \textit{StarkShift} and \textit{MultipoleMoments} modules in the JAC toolbox \cite{JAC}. This toolbox helps calculate (many-electron) interaction amplitudes of different kinds as well as the properties and a good number of excitation and decay processes for atoms and ions with complex shell structure. In some more detail, JAC provides (i) a rather intuitive user interface, (ii) a transparent data flow with and within the program, independent of the shell structure as well as (iii) features for modelling atomic cascades of different kinds \cite{atomic_cascades}.

All of these modules are available in JAC, which is hosted on GitHub \cite{JAC_GitHib}. Below, we provide an example of how JAC can be employed for a calculation of the QZSCs of $^3P_{0,1,2}$, $^1D_2$ and $^1S_0$ states of \ce{Ca^{14+}}. More specifically, this is a small script written in Julia, which imports JAC in the first line and can directly interact with the objects and functions defined within the toolbox. In this case, we don't directly interact with the module, handling the computation of the $C_2$ coefficient, but rather pass all of the necessary information by an \textit{Atomic.Computation} that handles all of the work and outputs a table containing all of the QZSCs. Of course, this example is purely for illustrative purposes and more information can be found in the documentation \cite{JAC_GitHib}.

% \begin{minted}[breaklines,escapeinside=||,mathescape=true, linenos, numbersep=3pt, gobble=2, frame=lines, fontsize=\small, framesep=2mm]{julia}
% using JAC
% grid    = Radial.Grid(true)
% nm      = Nuclear.Model(20.0, "Fermi")
% config  = Configuration("[He] 2s^2 2p^2")

% # Compute intermediate multiplet
% wa = Atomic.Computation(Atomic.Computation(), name="QZSC intermediate multiplet", grid=grid, nuclearModel=nm, configs=[config])
% wb = perform(wa, output=true)
% wbMultiplet = wb["multiplet:"]

% # Compute quadratic Zeeman shift coefficients
% lzSettings = LandeZeeman.Settings(false, false, false, true, true, true, 1.0, LevelSelection(), wbMultiplet)
% wc = Atomic.Computation(Atomic.Computation(), name="Quadratic Zeeman shift coeff.", grid=grid, nuclearModel=nm,
%     configs=[config],
%     propertySettings=[lzSettings])
% wd = perform(wc, output=true)
% \end{minted}

\begin{lstlisting}
using JAC
grid    = Radial.Grid(true)
nm      = Nuclear.Model(20.0, "Fermi")
config  = Configuration("[He] 2s^2 2p^2")

# Compute intermediate multiplet
wa = Atomic.Computation(
        Atomic.Computation(), 
        name="Intermediate multiplet",
        grid=grid, 
        nuclearModel=nm, 
        configs=[config])
wb = perform(wa, output=true)
wbMultiplet = wb["multiplet:"]

# Compute QZSCs
lzSettings = LandeZeeman.Settings(
            false, false, false, 
            true, true, true, 1.0, 
            LevelSelection(), 
            wbMultiplet)
wc = Atomic.Computation(
        Atomic.Computation(), 
        name="QZSCs",
        grid=grid,
        nuclearModel=nm,
        configs=[config],
        propertySettings=[lzSettings])
wd = perform(wc, output=true)
\end{lstlisting}

\bibliography{HCI_Shifts}

\end{document}